\documentclass[pra,twocolumn,showpacs]{revtex4}
\usepackage[colorlinks, linkcolor=blue,
urlcolor=blue]{hyperref}	
\usepackage{graphicx}
\usepackage{amsmath,amsfonts,amssymb,cases} 
\usepackage{cases}			
\DeclareMathOperator{\CPV}{P}

\newcommand{\dd}[1]{\frac{\mathrm{d}^d#1}{(2\pi)^d}}
\newcommand{\ddk}{\dd{k}}
\newcommand{\vc}[1]{\boldsymbol{#1}}
\renewcommand{\k}{{\vc{k}}}

\renewcommand{\r}{{\vc{r}}}
\newcommand{\dn}{\delta n}

\newcommand{\av}[1]{\overline{#1}} 
\newcommand{\lcor}{\sigma} 
\newcommand{\rmd}{\mathrm{d}} 

\begin{document}
\author{Christopher Gaul}
\author{Nina Renner}
\author{Cord A. M\"uller}
\title{Speed of sound in disordered Bose-Einstein condensates}
\pacs{03.75.Kk, %
63.50.-x
}

\affiliation{Physikalisches Institut, Universit\"at Bayreuth, D-95440
Bayreuth, Germany}

\begin{abstract}
Disorder modifies the sound-wave excitation spectrum 
of Bose-Einstein condensates. 
We consider the classical hydrodynamic limit, 
where the disorder correlation length is much longer
than the condensate healing length. 
By perturbation theory, we compute the phonon lifetime and the correction to the
speed of sound. 
This correction is found to be negative in all dimensions, with
universal asymptotics for smooth correlations. 
Considering in detail optical speckle potentials, we 
find a quite rich intermediate structure. 
This has consequences for the average density of states,
particularly in one dimension, where we find a ``boson dip'' next to a
sharp ``boson peak'' as function of frequency. 
In one dimension, our prediction is verified in detail by a numerical
integration of the Gross-Pitaevskii equation.  
\end{abstract}

\maketitle

\section{Introduction}

Disorder is a key feature for the understanding of the properties of matter.
In a disordered environment, waves 
can become coherently localized, leading to suppression of transport
 \cite{disoAnderson58,Billy2008}. Also strong
interaction can produce an insulator, via the Mott-Hubbard transition 
\cite{Greiner2002}.
The combined effects of interaction and disorder, despite being studied for decades, still
hold surprises. Here we are interested in the influence of a spatially
correlated disorder potential on the 
low-energy excitations of an interacting 
Bose-Einstein condensate (BEC). As Goldstone excitations, these
low-energy Bogoliubov excitations feature a
linear, phonon-like dispersion relation $\omega_k=ck$ with sound
velocity $c$. 
The sound velocity is of particular interest because it determines the
range of superfluidity, according to the Landau criterion, and it
determines the density of states, which enters virtually all
physically relevant quantities. 
Moreover, the speed of sound  is directly measurable in
cold-atom BECs \cite{Andrews1997}, where well-controlled 
optical speckle potentials with interesting spatial correlations
can be studied \cite{Billy2008,Lugan2009a}. 
 
Calculating the \emph{effective speed of sound} $\av{c}$ in disordered
systems is far from trivial, with different
approaches leading to different predictions. 
Perturbation theory, on the one hand, predicts an
\emph{increased} speed of sound due to 3D uncorrelated disorder
\cite{Giorgini1994,Falco2007}. 
On the other hand, within a self-consistent nonperturbative approach 
Yukalov and Graham 
\cite{Yukalov2007} reported numerically a slight \emph{decrease}. 
For disordered hard-core bosons on a lattice, 
Zhang \cite{Zhang1993} found a decrease of $c$ to fourth order in
disorder strength, without information on the second-order effect.

\begin{figure}
\includegraphics[width=0.9\linewidth]{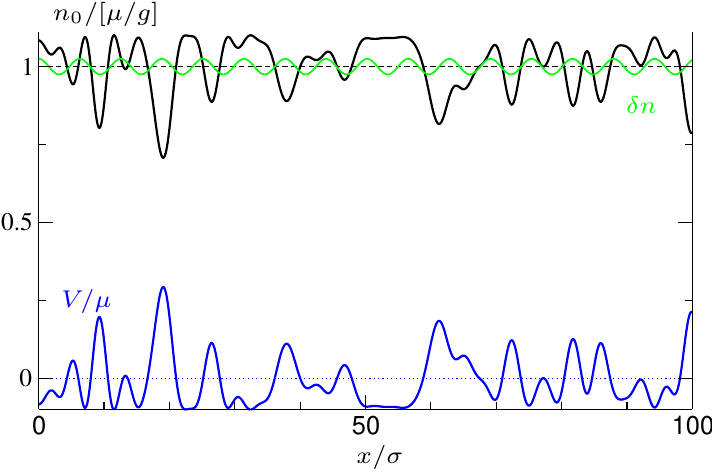}%
\caption{(Color online) 
Schematic 1d representation of the system under study: an interacting
Bose-Einstein condensate with original homogeneous density $n_0=\mu/g$
(dashed black line) 
is exposed to a weak, spatially correlated random potential $V(\r)$
(solid blue), here a blue-detuned speckle potential with amplitude 
$V=0.1\mu$, centered on the mean $\av{V}=0$ (dotted blue). 
We consider the Thomas-Fermi regime where the healing length
$\xi$ is much shorter than the disorder correlation length $\lcor$. 
The resulting ground-state density (solid
black) [eq.~\eqref{TFdensity.eq}]  mirrors
the disorder  while leaving the total average density and
number of particles constant. On top of this disorder-modified ground
state, an elementary plane-wave excitation
(green, plotted around 1) propagates with wave vector $k$, here with 
$k\lcor=1$. We calculate its effective speed of sound and the
corresponding average density of states.}
\label{setting.fig} 
\end{figure}

To clarify the situation 
with suitable parameters for present-day BEC experiments, 
we study in this article 
phonon excitations of a BEC in the strongly interacting
case where the chemical potential $\mu$ is much larger than the
disorder strength $V$, and where the 
condensate healing length $\xi =\hbar/\sqrt{2m\mu}$ is much smaller than
the disorder correlation length $\lcor$ 
(see Fig.~\ref{setting.fig}). Without disorder, the BEC is contained in a
very shallow trap and has a constant density
$n_0 = \mu/g$ in the region of interest. 
In the presence of smooth
disorder with correlation length $\lcor\gg\xi$, the
BEC ground state density follows the external potential with the
Thomas-Fermi profile $n_0(\r) =
n_0[1-V(\r)/\mu]$. 
A long-wavelength density deviation $\delta n(\r,t)$ from this ground state obeys
the wave equation  
\begin{equation}\label{wave.eq}
\left[\nabla \cdot c^2(\r) \nabla  -\partial_t^2 \right] \dn =0 
\end{equation}
where $c(\r) = c\left[1-V(\r)/\mu\right]^{1/2}$ is the local speed of
sound, deviating from the clean value 
$c = \sqrt{\mu/m}=\sqrt{gn_0/m}$. 
This is a prototypical wave equation in a 
medium with random elasticity, but constant mass density 
\cite{Gurarie2003,Gaul2007}. 
Quite often, the opposite case is
studied, with random masses and constant elasticity, or equivalently, a
fluctuating index of refraction
\cite{Gurarie2005,John1983,Akkermans2007}. The disorder potential 
may always be taken at zero average $\av{V(\r)}=0$. Its strength
is characterized by the variance $\av{V(\r)^2}=V^2$, and we suppose
weak disorder with $V\ll \mu$.

Consider now a sound wave with wave vector $k$ evolving on the
disordered potential background with correlation length $\lcor$ 
(see Fig.~\ref{setting.fig}). 
If the wavelength is much longer than the correlation length,
$k\lcor\ll1$, the 
excitation averages over the  potential fluctuations and, to a first
approximation, it seems reasonable to replace \eqref{wave.eq} by its
ensemble-average \cite{Gurarie2005}. But then we have no net effect on the speed of sound
since $\av{c^2(\r)} = c^2$, exactly. 
If, on the other hand, the wavelength is much shorter than the correlation length,
$k\lcor\gg1$, the excitation evolves in a locally constant potential, 
which should result in an average speed of sound 
$\av{c}=c \av{[1-V(\r)/\mu]^{1/2}} \approx c [1-\frac{1}{8}V^2/\mu^2]$. 
It turns out, however, that both these na\"{\i}ve reasonings fall
short. 

In order to give the correct answer right away, our main
results are briefly summarized in the following
section \ref{main.sec}. 
Section \ref{theory.sec} then presents the general
hydrodynamical perturbation theory, from which detailed results on the
speed of sound are derived in section \ref{sound.sec}. In section
\ref{dos.sec}, we analyze the implications of these results for the
disorder-averaged density of states. A short conclusion together with
a brief comparison to related
works are contained in section \ref{conclusion.sec}.

\section{Main results}
\label{main.sec}

The effective speed of sound in a disordered interacting Bose gas, properly defined as
$\omega_k/k=\av{c}$ from the single-excitation dispersion relation, is
affected by scattering processes via virtual intermediate states such
that a purely local description fails. We find that the correction 
$\Delta c = \av{c} - c$ to the speed of sound
has in $d$ dimensions the limiting behavior 
\begin{numcases} {\frac{\Delta c}{c} = -\frac{V^2}{2\mu^2} \times }
		d^{-1}, 	& $ k\lcor \ll 1$, \label{limits1.eq}\\ 
		\tfrac{1}{4}(2+d), 	& $ k\lcor \gg 1$. \label{limits2.eq}
\end{numcases}
These limits imply 
that the curves for different dimensions have to intersect around
$k\lcor \approx 1$ (see also Fig.~\ref{figRelSoundSpeckle} below). 
The precise shape of $\Delta c/c$ at intermediate $k\lcor$ 
depends on the details of the disorder pair correlation function.
But clearly, there is a negative correction, 
of order $V^2/\mu^2$, in all dimensions and for any disorder with finite
correlation length $\lcor \gg \xi$. 

A reduced speed of sound implies that 
the free density of states (DOS) of single excitations,
\begin{equation}
\rho_0(\omega)= \int\ddk \delta(\omega-ck)= \frac{S_d}{(2\pi
c)^d} \omega^{d-1},
\end{equation}
is replaced by an enhanced 
average density of
states (AVDOS) $\av{\rho}(\omega)$. Our results for this disorder-induced correction
can be cast into the form of a function 
\begin{equation}
g_d(\omega\lcor/c) =
\left[\av{\rho}(\omega) - \rho_0(\omega)\right] / \rho_0(\omega)
\end{equation} 
that depends only on the reduced momentum $\kappa
=\omega\sigma/c$: 
\begin{equation} \label{flimits.eq}
g_d(\kappa) = - \left[d+\kappa\frac{\rmd}{\rmd\kappa}\right] 
\frac{\Delta c}{c} = \frac{V^2}{2\mu^2} \times 
	\begin{cases} 
		1, 	& \kappa \ll 1, \\ 
		\frac{d}{4}(2+d), 	& \kappa \gg 1. 
	\end{cases}
\end{equation}
Gurarie and Altland \cite{Gurarie2005} suggested that one should be
able to deduce from the asymptotic values and the curvatures of such a
scaling function whether the AVDOS
exhibits a ``boson peak'' at intermediate frequency
$\omega \approx c/\lcor$.
The asymptotics of the scaling function in our case allow for a smooth, monotonic transition between the limiting values in any dimension $d$.
Thus one has no reason to expect any extrema in-between, which is indeed found to be the case in two and three dimensions.
In $d=1$ however, we find, by analytical calculation for the experimentally relevant case of an
optical speckle potential, a quite nonmonotonic AVDOS with
an intermediate dip followed by a sharp peak at
$\omega\sigma/c = 1$.

\section{Classical hydrodynamic theory}
\label{theory.sec}

We start our detailed analysis of the mean-field BEC order
parameter $\Psi = \sqrt{n} \, e^{i\phi}$
 in terms of the hydrodynamic variables 
condensate density $n = |\Psi|^2$ and phase $\phi$, which determines
the superfluid velocity $\vc{v} = \frac{\hbar}{m}\nabla
\phi$ \cite{Dalfovo1999,Pethick2002}. 
The grand-canonical Gross-Pitaevskii energy functional for the BEC in presence of an
external potential $V(\r)$ is 
\begin{align}\label{Efull.eq}
E [n,\phi] = \int \rmd^dr \biggl\lbrace  & 
\frac{\hbar^2}{2m}\left[ \bigl(\nabla \sqrt{n} \bigr)^2 + n(\nabla
\phi)^2\right]  \\ \nonumber &  
+ (V(\r)- \mu) n
+\frac{g}{2}n^2	\biggr\rbrace \ . 
\end{align}
The saddle-point equations $\delta E/\delta n|_0=0$ and $\delta E/\delta
\phi|_0=0$ imply that 
the ground state has constant phase $\phi_0$ or zero superfluid velocity
$\vc{v}_0 = 0$ and a density $n_0(\r)$ that obeys the stationary Gross-Pitaevskii equation
\begin{equation} \label{GPdensity.eq}
-\frac{\hbar^2}{2m}\frac{\nabla^2\sqrt{n_0(\r)}}{\sqrt{n_0(\r)}} + gn_0(\r)=\mu-V(\r).
\end{equation}

We now restrict our analysis to the case where the
healing length $\xi$ is much shorter than 
the disorder correlation
length $\lcor$. In this regime, the quantum pressure in
\eqref{GPdensity.eq} is
negligible, and the external disorder  potential leaves
the Thomas-Fermi imprint 
\begin{equation} \label{TFdensity.eq}
n_0(\r) = [\mu - V(\r)]/g.
\end{equation}
 This solution is also directly obtained if one drops the
density-gradient contribution in \eqref{Efull.eq} to use 
\begin{equation} \label{GPE2.eq}
E'[n,\phi] = \int \rmd^d r \left\{\frac{\hbar^2}{2m} n(\nabla\phi)^2 + [V(\r)-\mu]n +
\frac{g}{2}n^2 
\right\} . 
\end{equation} 
Formally, this formulation corresponds to the limit $\xi\to 0$, and all further
results can only depend on the reduced momentum $k\lcor$
\cite{Gaul2009a}.

The speed of sound characterizes the dynamics of small deviations
$\dn(\r,t) = n(\r,t)-n_0(\r)$ and $\delta \phi(\r,t) = 
\phi(\r,t)-\phi_0$ from the ground state in the long-wavelength regime
$k\xi\ll 1$. We can therefore develop the energy functional \eqref{GPE2.eq} to second
order around the ground-state solution, $E'=E_0'+F'[\delta n,\delta
\phi]$, to obtain the relevant quadratic energy functional 
\begin{equation} \label{F'.eq}
F'[\delta n,\delta \phi] = \frac{1}{2}\int \rmd^d r
\left\{\frac{\hbar^2}{m} n_0(\r)(\nabla\delta \phi)^2 + g\delta n^2 
\right\} .
\end{equation} 
Importantly, the external disorder potential has shifted the
ground-state solution according to \eqref{TFdensity.eq}, around which
we now consider the dynamics of fluctuations. Density and phase are
conjugate variables with the equations
of motion 
\begin{equation} 
\hbar\partial_t\delta n=\frac{\delta F'}{\delta(\delta\phi)},\quad 
-\hbar\partial_t\delta \phi=\frac{\delta F'}{\delta(\delta n)}. 
 \end{equation} 
In terms of density and superfluid
velocity, they read  
\begin{align} 
\partial_t\delta n+ \nabla \cdot [n_0(\r) \vc{v}] & = 0, \\
\partial_t \vc{v} = - \frac{g}{m}\nabla \delta n, 
 \end{align} 
and are recognized as the linearized versions of continuity
equation and Euler's equation for an ideal compressible fluid,
respectively. These can be combined to a single classical wave equation
\begin{equation}\label{waveV.eq}
\left[c^2 \nabla^2 -\partial_t^2 \right] \dn = \tfrac{1}{m}
\nabla \cdot \left [V(\r) \nabla \dn\right]. 
\end{equation}
This equation is equivalent to \eqref{wave.eq}, but now written in a
form amenable to systematic perturbation theory for a weak
external disorder potential $V(\r)$.

\subsection{Perturbation theory}

Translation invariance of the free equation suggests
to use a Fourier representation in space and time, 
\begin{equation}\label{eqWaveFourier}
 \left[\omega^2 - c^2 k^2\right] \dn_{\k} = 
 \int \dd{k'} \mathcal{V}_{\k\k'} \dn_{\k'}. 
\end{equation}
The disorder potential causes scattering $\k\to\k'$ of
plane waves with an amplitude 
\begin{equation}
\mathcal{V}_{\k\k'} = - \frac{1}{m}(\k \cdot \k')\,
V_{\k-\k'}.
\end{equation} 
The factor $\k \cdot \k'$ originates from the mixed gradient in
\eqref{waveV.eq} and implies pure $p$-wave scattering of sound waves
\cite{Gaul2008}  in contrast to $s$-wave scattering of independent
particles \cite{Kuhn2007}. 

The single-excitation dispersion relation 
can be derived from the corresponding Green function. 
The free Green function is diagonal in $\k$, 
\begin{equation}
G_0(k,\omega) = \left[\omega^2-c^2k^2+i0\right]^{-1}.
\end{equation} 
Taking the disorder average of the full Green function 
$G = \left[G_0^{-1} - \mathcal{V} \right]^{-1}$ 
leads in the standard way to \cite{Kuhn2007,Akkermans2007}
\begin{equation}
\av{G}(k,\omega) = \left[G_0(k,\omega)^{-1} - \Sigma(k,\omega)\right]^{\,-1}. 
\end{equation} 
The poles of this average Green function at 
$\omega^2 = c^2 k^2 + \Sigma(k,\omega)$ now determine the effective dispersion relation. 
The so-called self energy 
$\Sigma(k,\omega)$ is given to leading order in disorder strength
by the Born approximation:  
\begin{equation}\label{SigmaBorn}
\Sigma(k,\omega) = \int \dd{k'} \overline{[\mathcal{VV}]}_{\k\k'}
G_0(k',\omega). 
\end{equation}
The scattering potential correlator 
\begin{equation}
\label{VVcorr.eq}
 \overline{[\mathcal{VV}]}_{\k\k'} =
m^{-2}V^2\left[\k \cdot \k' \right]^2 \sigma^d P_d(|\k'-\k|\lcor) 
\end{equation}
involves the dimensionless $k$-space 
correlator of the bare potential
\begin{equation}
 P_d(\kappa) = \int {\rm d}^d\rho \, e^{-i\vc{\kappa}\cdot\vc{\rho}} C_d(\rho) .
\end{equation}
Its real-space correlator 
$C_d(r/\lcor) = \overline{V(\r)V(0)}\, / \,{V}^2$ is assumed to be
isotropic. We will consider correlated potentials for which 
$C(r/\lcor)$ decays from $C_d(0)=1$ to 0 on the length scale
$\lcor$. The smoothness of $V(\r)$ implies that the power spectrum $P_d(\kappa)$ 
decreases rapidly to 0 as function of $\kappa=k\lcor$.

Applying Sokhotsky's formula $(x + i\,0)^{-1} = \CPV
\frac{1}{x} - i \pi \delta(x)$ to the free Green function in
\eqref{SigmaBorn}, we can evaluate the real part and the imaginary
part of the self-energy separately. The imaginary part determines the
lifetime $\gamma^{-1}$ of the 
excitations, whereas its real part shifts the 
speed of sound by $\Delta c$: 
\begin{equation}\label{Deltagamma}
\frac{\Sigma(k,c k)}{2 c^2 k^2} = \frac{\Delta c}{c} - i \,
\frac{\gamma}{2ck}. 
\end{equation} 
To leading order in $V$, the on-shell dispersion $\omega=ck$ 
is used for evaluating the self-energy.

\subsection{Scattering rate and 1d localization length}

Calculating the imaginary part in \eqref{SigmaBorn}, 
the scattering rate at frequency $\omega=ck$ can be
expressed as 
\begin{equation}\label{eqElastScatt}
\gamma(\omega) = \frac{\pi V^2}{2\mu^2} \omega^2 \rho_0(\omega)\sigma^d
f_d\left(\omega\sigma/c\right) . 
\end{equation}
The last factor is the angular average of the correlation
function on the energy shell: 
\begin{equation}
\label{fdofu.eq}
f_d(\kappa) = S_d^{-1}\int \rmd\Omega_d (\cos\theta)^2
P_d(2 \kappa \sin\frac{\theta}{2}).
\end{equation} 
The squared cosine 
under the integral goes back to the $\left[\k \cdot \k' \right]^2$ in
the potential correlator \eqref{VVcorr.eq}, being again characteristic for $p$-wave scattering of sound
waves.

In one dimension, 
there are only the two contributions $\theta=0,\pi$ of forward- and
backscattering, respectively, such that 
\begin{equation}
\label{gamma1d.eq} 
\gamma(\omega) = \frac{V^2\sigma}{4\mu^2c} \omega^2
\left[P_1(0)+P_1(2\omega\sigma/c)\right].
\end{equation}

We note in passing that the 1d backscattering process $k \mapsto -k$ described by the second
contribution $P_1(2\kappa)$ is known to induce strong, Anderson localization of
the excitation in the disordered potential
\cite{KramerMacKinnon1993}. The backscattering rate is
directly proportional to the inverse localization length 
$\Gamma_\text{loc} = \gamma_\text{bs}/2c_0$ describing exponential localization 
\cite{Thouless1973}. Taking the backscattering contribution of
\eqref{gamma1d.eq}, we find  
\begin{equation}
 \Gamma_\text{loc} 
 = \lcor \frac{V^2}{8\mu^2} \, \frac{\omega^2}{c^2} \, P_1(2 k\lcor), 
\end{equation}
which agrees with the findings of a hydrodynamic theory similar to
ours \cite{Bilas2006} and also with the sound-wave limit of 
Bogoliubov excitations considered in \cite{Lugan2007a}. It should be
noted that these
latter approaches employ the phase-formalism that is particularly
suited for 1d systems, whereas our Green-function theory permits to go
to higher dimensions without conceptual difficulties. 

In any dimension,  the phase function $f_d(\kappa)$ in \eqref{fdofu.eq}
tends to a constant for small $\kappa$, and the
scattering rate of low-energy excitations 
rate tends to zero as 
$\gamma/\omega \propto \omega^{d}$. 
Also the localization length in 1d
diverges as $\Gamma_\text{loc}^{-1} \propto \omega^{-2}$ at low
frequency. In higher dimensions, it is known to be even larger, if not
infinite \cite{John1983}.
This 
assures that low-energy excitations are long-lived and extended. It is
thus meaningful to discuss their effective sound velocity. 

\section{Effective speed of sound}
\label{sound.sec}

From \eqref{SigmaBorn} and \eqref{Deltagamma}, the
speed-of-sound shift $\Delta c$ in any dimension $d$ is 
obtained as a Cauchy principal value integral
over the potential correlation:
\begin{equation} 
\frac{\Delta c}{c}
= -\frac{1}{2} \frac{V^2}{\mu^2} \CPV
\int \dd{(k'\lcor)} \, \frac{\left[\k \cdot \k'
\right]^2P_d(|\k'-\k|\lcor)}{k^2\, (k'^2 - k^2)} .
\label{Deltac.eq}
\end{equation}
In the limits $k\lcor\gg1$ and $k\lcor\ll1$ 
where the potential appears very smooth or $\delta$-correlated,
respectively, over a wavelength of the propagating excitation, this
correction is independent of the precise form of the bare potential correlator
$P_d(\kappa)$ (see \eqref{limits1.eq} and \eqref{limits2.eq} above and the detailed derivation in
section \ref{higherd.sec} below). Let us then 
discuss the interesting, detailed form of this 
correction as function of $\kappa=k\lcor$ in $d=1$.
For concreteness, we study the
case of an optical speckle potential, which has recently been
successfully used in experiments on Anderson localization of matter waves
\cite{Billy2008}. 

\subsection{Speckle potential} 

By focusing a laser beam through a diffusor, the condensate is subject
to a random lightshift potential proportional to the intensity of the laser
field \cite{Clement2006}. 
The one-point potential value $V(\r)$ 
of a speckle pattern \cite{Goodman1975} 
has the skewed probability
distribution 
\begin{equation} \label{Pdistribution.eq}
P(w)\rmd w=\Theta(1+w)\exp[-(1+w)]\rmd w, 
\end{equation}
for $w=V(\r)/V$. For this one-sided exponential, 
odd moments, such as $\av{V(\r)^3}$, are different from zero. 
A blue-detuned
light-shift potential with $V>0$ features repulsive peaks
(this case is depicted in Fig.~\ref{setting.fig}),
whereas a red-detuned one with $V<0$ consists of attractive wells.
 As far as spatial correlations are concerned, the laws of optics forbid
variations on a length scale shorter than the correlation length
$\lcor$, which depends on the laser wavelength and the geometry of the
imaging system, but typically ranges around $1\,\mu$m.  
In one dimension, the correlation function is 
\begin{equation}\label{c1.eq} 
P_1(\kappa) =  \frac{\pi}{2} (2-|\kappa|)\Theta(2-|\kappa|) . 
\end{equation}
Its bounded support in $k$-space implies that within the Born
approximation, backscattering and inverse localization length vanish 
for $k\lcor>1$; however, exponential localization still prevails due to higher orders
in perturbation theory \cite{Lugan2009a}.

\subsection{In dimension $d=1$}

\begin{figure}
\includegraphics[width=\linewidth]{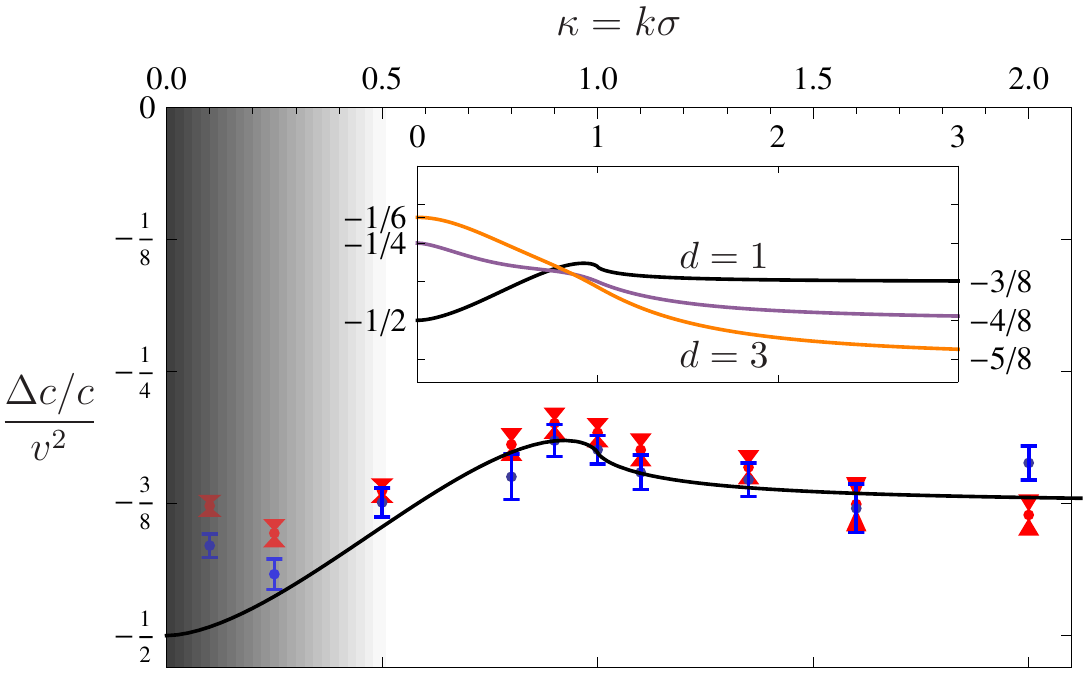}%
\caption{(Color online) Relative correction to the speed of sound as function of
reduced momentum. Black line: analytical prediction \eqref{eqSpeckle1D}
for $d=1$ in the limit $\xi/\lcor=0$. 
Blue/red symbols: data from a numerical integration of the 
Gross-Pitaevskii equation (with small but finite healing length, such that $k\xi
= 0.05$) at $v=V/\mu=\pm 0.03$, respectively, 
averaged over 50 realizations of disorder. 
In the shaded area, the data strays far from the analytical prediction  
because its condition of applicability $\sigma\gg\xi$ becomes invalid.  
Inset: Theory \eqref{Deltac.eq} for $d=1$ 
(black), $d=2$ (violet), and $d=3$ (orange) together with the limiting values \eqref{limits1.eq} and
\eqref{limits2.eq}.}
\label{figRelSoundSpeckle} 
\end{figure}

The principal-value integral \eqref{Deltac.eq} over the piecewise
linear function \eqref{c1.eq}
is elementary, and we find a speed-of-sound correction
at $\kappa=k\lcor$ of 
\begin{equation}\label{eqSpeckle1D}
\frac{\Delta c}{c} 
= - \frac{V^2}{2\mu^2}
\left( 1 + \frac{\kappa	}{4}\ln\left|\frac{1-\kappa  }{1+\kappa}  \right|
               - \frac{\kappa^2	}{4}\ln\left|\frac{1-\kappa^2}{\kappa^2}\right|
\right) . 
\end{equation}
Its limiting values are 
${\Delta c}/{c} = -\frac{1}{2}{V^2}/{\mu^2}$ for small $\kappa$, and
${\Delta c}/{c} = -\frac{3}{8}{V^2}/{\mu^2}$ for large $\kappa$, 
as stated in \eqref{limits1.eq} and \eqref{limits2.eq}. This correction
to the speed of sound, plotted in Fig.\
\ref{figRelSoundSpeckle} as function of  $\kappa=k\lcor$, shows a
rather intricate, non-monotonic behavior. 
Notably, there is a logarithmic non-analyticity at $\kappa = 1$, the
value beyond which backscattering is suppressed. The
speed-of-sound correction is clearly \emph{negative} for all $\kappa$,
which may come as a surprise in view of
\cite{Giorgini1994,Falco2007}. 

In order to check this prediction in detail, we have numerically
integrated 
the full Gross-Pitaevskii equation describing
an elementary excitation with fixed $k$ on top of the numerically
determined groundstate in a speckle potential with $V=\pm 0.03\mu$
and variable $\sigma$. 
This simulation operates at a small but finite
value of $k\xi = 0.05$ and includes the full quantum
pressure. Moreover, it does not rely on a linearization for small 
excitations nor perturbation theory in $V$. We extract the effective
dispersion $\omega_k$ 
by monitoring the phase of $\Psi_k(t)$ and then find
$\av{c}=\omega_k/k$. As shown in Fig.~\ref{figRelSoundSpeckle}, the
data agree beautifully with
\eqref{eqSpeckle1D} in its realm of validity, $\lcor\gg \xi$. 
When the correlation length decreases towards the healing
length (shaded area in
Fig.~\ref{figRelSoundSpeckle}), the
correction vanishes at fixed disorder strength, because the condensate
density is smoothed with respect to the Thomas-Fermi profile
\cite{sanchez-palencia06}. 
But in any case, only negative corrections are found in $d=1$.

\subsection{In higher dimensions} 
\label{higherd.sec}

In higher dimensions, the integral \eqref{Deltac.eq} 
over the correlation functions (for speckle, see \cite{Kuhn2007})
is sufficiently complicated  that analytical solutions like \eqref{eqSpeckle1D} are not available in general.
(As an exception to this rule, we find  for the 2D speckle
correlation $\Delta c/c = -\frac{V^2}{8\mu^2}(4-\kappa^{-2})$ for $\kappa>1$.)
But in all cases, the principle-value integral \eqref{Deltac.eq} can be evaluated
numerically. The inset of Fig.~\ref{figRelSoundSpeckle} shows the
corresponding curves.
Short-range correlated potentials ($k\lcor\ll1$) affect low dimensions more than high dimensions and vice versa.

The limits \eqref{limits1.eq} and \eqref{limits2.eq} 
can be calculated analytically as follows. 
It is useful to
rewrite \eqref{Deltac.eq} in terms of $\eta=1/k\sigma$ as 
\begin{equation}
\frac{\Delta c}{c} 
= -\frac{V^2}{2\mu^2} 
\CPV \int \dd{q}   \frac{P_d(q)\left[1+ \eta q\cos{\beta}\right]^2}{2 \eta
q \cos{\beta} +\eta^2 q^2} . \label{eqSigmaRel2}
\end{equation}
Denoting the angular part of the integral by $A_d(\eta q)$, one arrives at the radial integral
$\int_0^\infty {\rm d} q\, q^{d-1}  P_d(q) A_d(\eta q)$. 
In the limit $k\lcor \ll 1$, the parameter $\eta q$ tends to
infinity nearly everywhere under the integral. Then 
\begin{equation}
A_d(\infty) = \int \frac{{\rm d}\Omega_d}{(2\pi)^d} (\cos\beta)^2 =
\frac{S_d}{(2\pi)^{d}}d^{-1},
\end{equation}  
and with $\int \dd{q} P_d(q) =C_d(0)= 1$, we arrive at \eqref{limits1.eq}.
In the limit $k\lcor \to\infty$ we proceed similarly with $\eta \to 0$. 
The angular integrand reduces to $1 + \left[ {2}\eta \cos\beta +
\eta^2\right]^{-1}$, whose principle-value integral evaluates after some algebra
to 
\begin{equation}
A_d(0) = \frac{S_d}{(2\pi)^d}\frac{d+2}{4},
\end{equation} 
which leads to \eqref{limits2.eq}.

\subsection{Numerical investigation beyond Born}

\begin{figure}
\includegraphics[width=\linewidth]{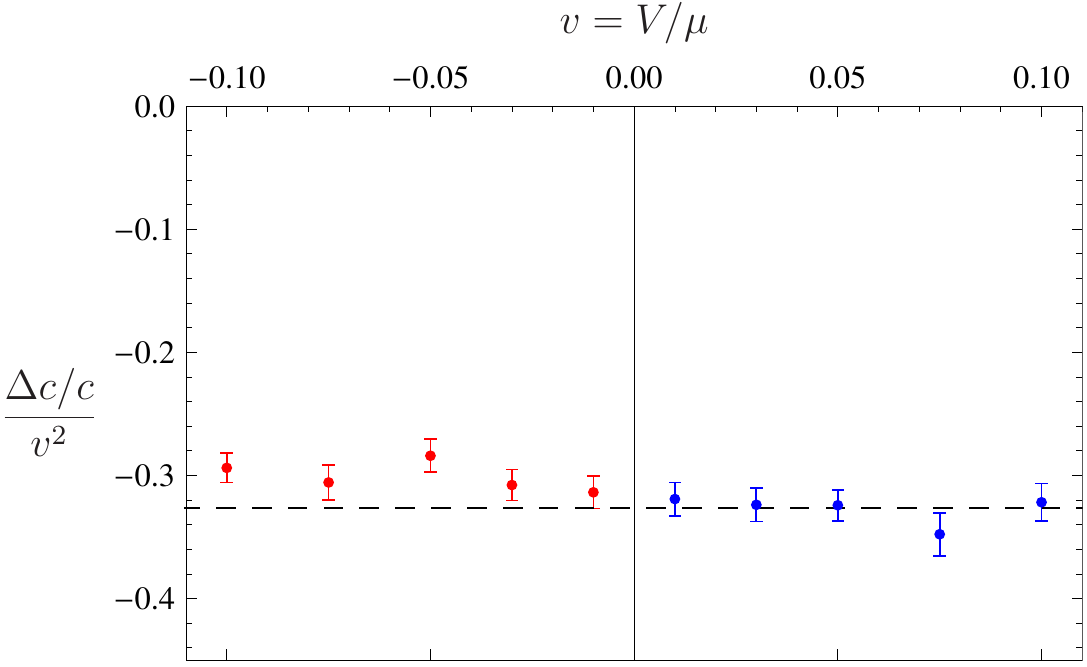}
\caption{(Color online) 
Relative correction to the speed of sound at $k\lcor = 1$, divided by
$v^2$,  
as function of disorder strength $v=V/\mu$, taken 
from the numerical integration with $k\xi=0.05$.   
Dashed line: analytical prediction $\Delta
c/(c v^2)= \frac{1}{4}\ln 2-\frac{1}{2} \approx -0.3267$.}
\label{figV}
\end{figure}

The perturbative analysis relies
on the Born approximation \eqref{SigmaBorn}, so that good agreement
with the true values is only expected at rather small disorder. Could a larger
disorder strength reverse the sign of the correction? 
We have numerically investigated different values of $v=V/\mu$ at
fixed $k\lcor=1$. In 
Fig.~\ref{figV}, we show the data divided by $v^2$ such that the Born
approximation shows as a horizontal line. 
As expected, for small $|v|$ the agreement is very satisfactory. 
One can distinguish a third-order correction $O(v^3)$ as a linear 
trend with negative slope; if needed, it could be calculated pushing
\eqref{SigmaBorn} beyond the Born approximation \cite{Lugan2009a}. 
In a Gaussian
model with a symmetric probability distribution \cite{Hartung2008},
such a third-order term would be absent. 

\begin{figure}
\includegraphics[width=0.65\linewidth,angle=270]{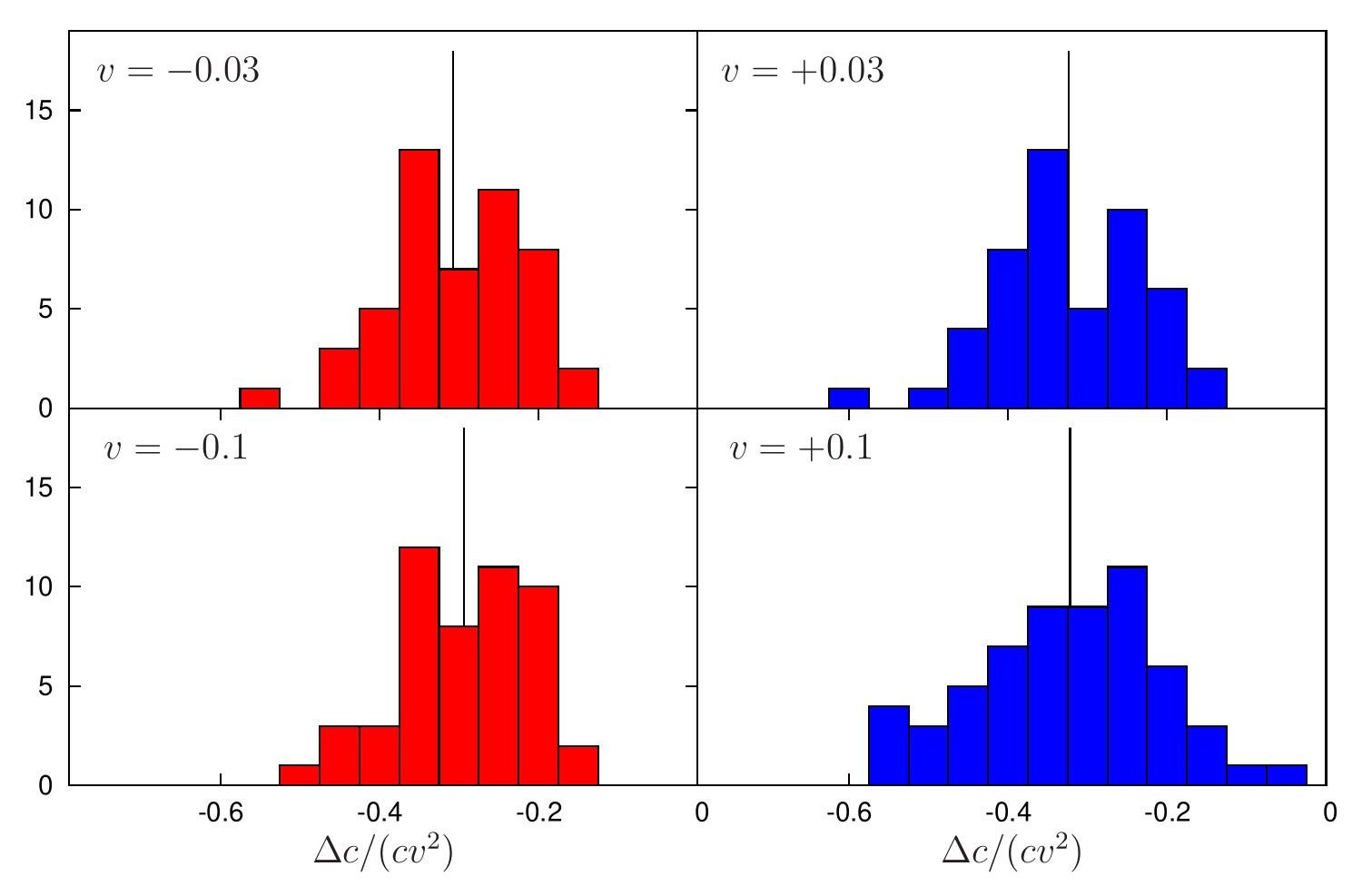}%
\caption{(Color online) 
Histograms over 50 realizations of disorder 
for the relative speed-of-sound correction for different values of $v=V/\mu$ 
obtained by numerical integration of the 
Gross-Pitaevskii equation with an excitation at $k\lcor=1$.   
Vertical lines indicate the average
values plotted in Fig.~\ref{figV}.}
\label{histogram.fig}
\end{figure}

The error bars in Figs.~(\ref{figRelSoundSpeckle}) and (\ref{figV}) indicate the
estimated error of the mean after ensemble-averaging 
over 50 realizations of disorder. 
Fig.~\ref{histogram.fig} displays exemplary histograms of the
values obtained for different disorder realizations. Clearly, the probability distributions are
single-peaked with well-defined averages on the negative side. 
That the speed of sound has self-averaging character was to be 
expected since a plane wave samples different spatial regions
at once. At strong disorder with $v \gtrsim 0.1$, the speckle disorder
with its unbounded probability distribution is likely to fragment
the condensate, and the concept of a unique, well-defined speed of sound
becomes questionable.  As a precursor, we already observe a slight broadening of
the probability distribution for $v=+0.1$ (lower right panel).

From the data shown, we conclude that the
correction to the speed of sound remains \emph{negative} over the
entire interval of interest.

\section{Density of states} 
\label{dos.sec}

\begin{figure}
\includegraphics[width=\linewidth]{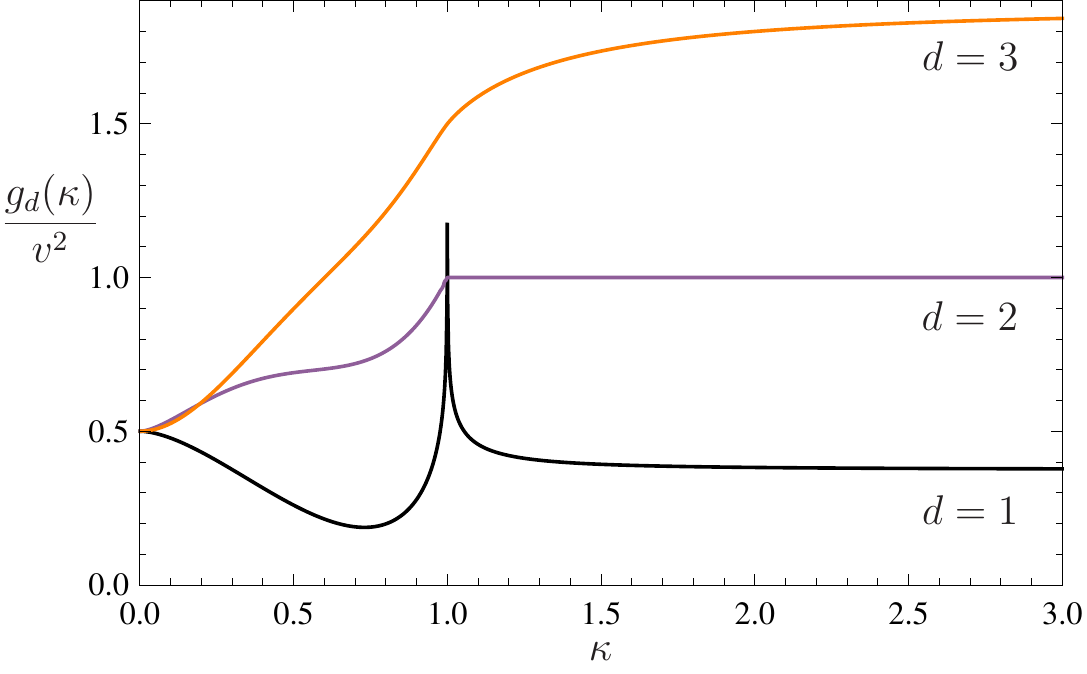}
\caption{Correction to the density of states $g_d(\kappa) =
(\overline \rho -\rho_0)/\rho_0$ divided by the squared disorder strength
$v=V/\mu$ as function of reduced momentum 
$\kappa = \omega \sigma/c$ in dimension $d=1,2,3$. 
At $\kappa = 1$, the momentum beyond which elastic
backscattering becomes impossible in the Born approximation, there is a
logarithmic divergence in $d=1$, a kink in $d=2$, and a curvature
discontinuity in $d=3$.}\label{figDOS} 
\end{figure}

Knowing the speed of sound, we can compute the average density of
states (AVDOS) 
\begin{equation} 
\av{\rho}(\omega)= \int\ddk \delta(\omega-\omega_k)
\end{equation}
 using the effective dispersion $\omega_k=\av{c}(k)k$ 
in the perturbative limit where $\gamma/\omega\ll 1$. 
Denoting, similarly to \cite{Gurarie2005},
\begin{equation}
\av{\rho}(\omega)= \rho_0(\omega)
\left[1+g_d(\omega\lcor/c)\right],
\end{equation} 
we find for the relative correction
\begin{equation} 
g_d(\kappa) = - \left[d+\kappa\frac{\rmd}{\rmd\kappa}\right] 
\frac{\Delta c}{c} 
\end{equation}
with the limiting values \eqref{flimits.eq} 
\cite{remarkDOSBilas}. 
The scaling functions
$g_d(\kappa)$ are plotted in Fig.~\ref{figDOS} for dimension $d=1,2,3$. 
In one dimension, 
\begin{equation}
g_1(\kappa) = \frac{v^2}{2}
\left(1 + \frac{\kappa}{2}  \ln\left|\frac{\kappa-1}{\kappa+1}\right|  
	- \frac{3\kappa^2}{4} \ln\left|\frac{1-\kappa^2}{\kappa^2}\right| 
\right)
\end{equation}
shows a pronounced dip around
$\kappa \approx 0.7$ and a sharp logarithmic divergence at
$\kappa=1$. This particular structure is a consequence of the Born approximation,
more specifically the non-analyticity of the speckle
pair correlation function at the boundary of its support. But a
local maximum is also found for other correlated potentials with fast enough
decay of $P_d(q)$ such as Gaussian correlation \cite{Hartung2008}. 
The existence of this  ``boson moat'' could not
be inferred from the asymptotics of $g_1(\kappa)$ alone \cite{Gurarie2005}. Indeed, 
expanding the asymptotic behavior as 
\begin{equation}
g_d(\kappa) = v^2\times 
	\begin{cases} 
		\beta_d^<(1+\alpha_d^<\kappa^2+\dots ), 	& \kappa \ll 1, \\ 
		\beta_d^>(1+\alpha_d^>\kappa^{-2}+\dots ), 	&
\kappa \gg 1, 
	\end{cases}
\end{equation}
we find $\alpha_1^<= -1-\frac{3}{2}|\ln\kappa| <0$ and
$\alpha_1^>=\frac{1}{18} > 0$ of opposite sign. 
Together with the fact that $ \beta_1^<$ is larger than $\beta_1^>$,
these asymptotics would be compatible with a monotonic
behavior and thus are not sufficient to infer the existence of
intermediate extrema. 

In two dimensions, 
the scaling function is exactly constant for $\kappa >1$ and thus
$\alpha_2^>=0$ which seems to happen also in other cases
\cite{Gurarie2005}. At $\kappa = 1$, there is a kink,
but overall, $g_2(\kappa)$ shows a monotonic behavior without local
extrema.  
In three dimensions, the logarithmic singularity has moved to the 
second derivative of $g_3(\kappa)$, which is hardly resolvable in the
figure, and $\alpha_3^><0$ as expected \cite{Gurarie2005}, leaving an
all but structureless AVDOS.

\section{Conclusions}
\label{conclusion.sec}

In conclusion, we have perturbatively calculated the influence of a weak spatially
correlated disorder potential on the sound-wave spectrum 
of Bose-Einstein condensates  
in the hydrodynamic limit $\xi \ll \lcor, 1/k$ and arbitrary
dimension. The sound-wave 
lifetimes are long enough to observe a disorder-induced correction to
the speed of sound, which is found to be \emph{reduced}. 
For the experimentally relevant case of an optical speckle potential, 
we compute the correction to the speed of sound analytically. 
A numerical integration of the full mean-field dynamics 
in $d=1$ confirms our prediction in its range of validity and even 
allows to access non-perturbative disorder strengths.  

The present hydrodynamic theory compares well to
results in $d=1$ from the phase-formalism approaches of Bilas and
Pavloff \cite{Bilas2006} and Lugan et al.~\cite{Lugan2007a}. We find
perfect agreement concerning the localization length, which we obtain
from the backscattering rate.

However, our results are in contrast to the impact of uncorrelated disorder in three dimensions,
for which Giorgini, Pitaevskii and Stringari \cite{Giorgini1994} have 
predicted a positive correction to the speed of sound. 
Yet, at present there appears no contradiction between their results and
ours. We have used a simple hydrodynamic
description valid for $\xi\ll\lcor,1/k$ that
cannot cover the case of a truly $\delta$-correlated disorder, spatially
varying on a scale $\lcor\ll\xi$, considered by Giorgini et al. 
In particular, for such rapidly varying potentials  the Thomas-Fermi approximation
Eq.~\eqref{TFdensity.eq} for the ground state density does not hold
any more and should be replaced by the solution of
\eqref{GPdensity.eq}, which then shows the smoothed imprint of the disorder
potential \cite{sanchez-palencia06}.
      
Furthermore, we have determined the average sound-wave 
density of states. In low dimensions, its structure is very rich,
including a ``boson moat'' consisting of a broad dip followed by a
sharp peak in $d=1$. As a rule, specific correlation-related 
features tend to be washed out by integration in higher dimensional
$k$-space. Thus we
expect arguments on general grounds \cite{Gurarie2005} to hold more
reliably in higher dimensions. Conversely, the low-dimensional
behavior may escape a bird's-eye view  and require detailed
calculations. We have presented such a calculation for spatially
correlated speckle disorder, so that our results should be of
immediate use for cold-atom experiments.  

\begin{acknowledgments}

This research was supported by DFG and DAAD. 
We thank P.~Bouyer and L.~Sanchez-Palencia for generous hospitality at
Institut d'Optique, Palaiseau, and
helpful discussions.
\end{acknowledgments}


\begin{thebibliography}{10}
\bibitem{Gaul2009a}
The full quantum problem with three independent length scales
$\lambda,\xi,\lcor$ requires a more
  elaborate formulation that will be published elsewhere.
\end{thebibliography}
\end{document}